\documentclass[preprint,12pt]{aastex}
\usepackage{xfrac}
\usepackage[caption=false]{subfig}
\usepackage{hyperref} 
\hypersetup{
	breaklinks,
	colorlinks=true,
	pdfauthor={Charles J. Lada},
	pdftitle={On the Mass-Size Relation for GMCs},
	citecolor=blue
	}

\usepackage{longtable}

\newcommand{\msun}{M$_\odot$}

\newcommand{\twco}{$^{12}$CO\ }

\newcommand{\sdu}{M$_\odot$ pc$^{-2}$}
\newcommand{\sdusp}{M$_\odot$ pc$^{-2}$\ }
\newcommand{\sd}{$\Sigma_{GMC}$}
\newcommand{\sdsp}{$\Sigma_{GMC}$\ }
\newcommand{\xco}{X$_{CO}$\ }
\newcommand{\md}{MD+17\ }
\newcommand{\rice}{R+16\ }

\newcommand{\rgal}{R$_{\rm gal}$}

\shorttitle{Protostars}
\shortauthors{Lada \& Dame}
\begin{document}
\title{ The Mass-Size Relation and the Constancy of GMC Surface Densities in the Milky Way}
\author{Charles J. Lada \& T. M. Dame}
\affil{Harvard-Smithsonian Center for Astrophysics \\ 60 Garden St \\Cambridge, MA 02138}
\email{clada@cfa.harvard.edu} 
\email{tdame@cfa.harvard.edu}

\begin{abstract}

We use two existing molecular cloud catalogs derived from the same CO survey and two catalogs derived from local dust extinction surveys to investigate the nature of the GMC mass-size relation in the Galaxy.  We find that the four surveys are well described by  $M_{GMC} \sim R^2$ implying a constant mean surface density, $\Sigma_{GMC}$, for the cataloged clouds. However, the scaling coefficients and scatter differ significantly between the CO and extinction derived relations. We find that the additional scatter seen in the CO relations is due to a systematic variation in $\Sigma_{GMC}$\   with Galactic radius that is unobservable in the local extinction data. We decompose this radial variation of $\Sigma_{GMC}$\  into two components, a  linear negative gradient with Galactic radius and a broad peak coincident with the molecular ring and superposed on the linear gradient. We show that the former may be due to a radial dependence of X$_{CO}$ on metallicity while the latter likely results from a combination of increased surface densities of individual GMCs and a systematic upward bias in the measurements of $\Sigma_{GMC}$\  due to cloud blending in the molecular ring. We attribute the difference in scaling coefficients between the CO and extinction data to an underestimate of X$_{CO}$. We recalibrate the CO observations of nearby GMCs using extinction measurements to find that locally X$_{CO}$ $=$ 3.6$\pm$0.3 $\times$ 10$^{20}$ cm$^{-2}$ (K-km/s)$^{-1}$.  We conclude that outside the molecular ring the GMC population of the Galaxy can be described to relatively good precision  by a constant $\Sigma_{GMC}$\  of 35 M$_\odot$ pc$^{-2}$.

\end{abstract}

\maketitle

\section{Introduction}
Giant molecular clouds (GMCs) play a pivotal role in star formation and galaxy evolution. 
Stars form from such cold and massive clouds at almost every epoch of cosmic evolution. Deciphering the physical nature and evolution of GMCs is a necessary step for understanding the process of star formation and ultimately galaxy evolution.  Within a decade of their discovery, Larson (1981) compiled existing CO observations of nearby GMCs and identified three basic empirical scaling relations obeyed by these objects: 1) a power-law scaling between velocity dispersion and cloud size, i.e., $\sigma_v \sim R^{0.5}$, 2) an approximate state of virial equilibrium for the clouds,  i.e.,  $5\sigma_v^2R/GM = 1$, and 3) a power-law scaling between GMC mass and size, i.e., $M_{GMC} \sim R^2$. The last relation implies that GMCs have constant average column densities. The scatter in all these relations is typically large (0.4 - 0.5 dex) (e.g., Larson 1981, Solomon et al. 1987, Falgarone et al. 2009) raising the question of how precisely do the Larson relations describe the nature of GMC populations  in the Milky Way. In other words, is the observed scatter in these relations largely due to experimental uncertainty inherent in the CO observations, or is much of the scatter intrinsically physical? In the first instance, individual GMCs in a population would closely conform to these relations (e.g., the GMCs would all have a very similar surface density), while in the second instance the Larson relations are obeyed only in some average sense for the GMC population. 

Using dust extinction rather than CO to measure GMC masses and sizes, Lombardi et al. (2010) re-examined the GMC mass-size relation for a local sample of GMCs and found an extremely tight power-law scaling between these two quantities. The power-law index was found to be 2 with a measured scatter of only 11\% or 0.04 dex. This scatter is significantly  lower than found for any of the Larson scaling relations using CO data. Indeed, Lombardi et al.'s result indicates that local GMCs are characterized to high degree of accuracy by a constant average column or mass surface density which was directly measured to be  $\Sigma_{GMC} = $ 41 $\pm$ 5 M$_\odot$ pc$^{-2}$.   Ballesteros-Paredes et al. (2012) showed that the measurement of such a precisely constant surface density for GMCs was a natural consequence of the facts that GMCs have power-law column density pdfs that decrease relatively steeply with column density (Lombardi et al. 2015) and that the average column density is systematically computed for gas lying above a fixed column density threshold (typically corresponding to A$_V$ $\approx$ 1 mag.).  The latter essentially guarantees that the computed average column density will be within some small factor of the value of the adopted threshold density. Indeed,  Lombardi et al. (2010) found this factor to be $\sim$ 2  for the clouds in their sample. 
Consequently, to the extent that GMCs in a given population have similar structure and their masses and sizes are systematically measured from the same threshold column density, their computed average surface densities should be always nearly the same. One interesting consequence of this finding is that local GMCs cannot obey a Kennicutt-Schmidt star formation scaling law (Lada et al. 2013). 

Why is it then that reported CO measurements of cloud surface densities in the Milky Way are in the range $\Sigma_{GMC}$ $\sim$ 2 - 400 M$_\odot$ pc$^{-2}$, or equivalently, A$_V$ $\sim$ 0.1 - 180 magnitudes (e.g., Solomon et al. 1987; Roman-Duval et al. 2010, Heyer and Dame 2015, Miville-Deschenes et al. 2017)? Such measurements clearly contradict both Larson's and Ballesteros-Paredes et al.'s constant column density predictions. Possible explanations of this dilemma include: 1)- uncertainties in the determinations of cloud sizes and masses from CO data. Such uncertainties could arise from effects such as the definition of a cloud, the use of variable surface density thresholds to define cloud boundaries, the disentangling of cloud overlap along the line-of-sight,  variation in the CO mass conversion (\xco) factor, etc., 2)-  the scaling coefficient that characterizes the mass-size relation is  variable and 3)- some combination of 1) and 2). 

In this paper we re-examine the mass-size relation for GMCs in the Milky Way. We use four different GMC catalogs which employed different tracers of molecular material and differing methodologies to identify and extract the physical properties of the clouds. Two of these catalogs used CO observations to trace the molecular gas and two used observations of dust extinction for the same purpose. 

\section{\label{results} Data}

The \twco data were drawn from the recent molecular cloud catalogs of Rice et al. (2016; hereafter R+16) and Miville-Deschenes et al (2017; hereafter MD+17). Both these catalogs used the complete Galaxy-wide $^{12}$CO survey by Dame et al. (2001; hereafter DHT) to identify and measure the basic properties of  GMCs across the Galaxy.  The  DHT survey is a composite of large scale CO surveys obtained with a pair of 1.2 m telescopes covering the Northern and Southern skies with an angular resolution of 8.5 arc min. It is the most complete and uniform CO survey of the Milky Way yet produced. 

R+16 performed a dendrogram-based decomposition of the DHT survey to identify 1064 GMCs with masses ranging from $\sim$ 2.5 $\times$ 10$^3$ to 10$^7$ \msun\ and sizes (radii) between $\sim$ 2-240 pc, in total recovering about 40\% of the CO emission in the DHT survey. Distances to the clouds were derived from the Galactic rotation curve. For the present study we omitted a small number of clouds for which reliable distances could not be determined, resulting in a cleaned catalog containing 1037 GMCs. MD+17 employed a gaussian decovolution followed by a cluster finding algorithm to recover 8,107 molecular clouds accounting for more than 98\% of the CO emission in the DHT survey.  For the present study we filtered the MD+17 catalog by removing sources that, on independent inspection of the DHT survey, did not appear real. To identify the spurious sources we employed the smooth-masking technique described by Dame (2011) to isolate regions of the sky with significant CO emission from those with no detectable CO emission. We then located all the sources in the MD+17 catalog whose positions coincided with regions of null CO emission in the smoothed masks and removed them from the catalog. This resulted in a cleaned catalog containing 5577 sources with confident identifications. The sources that were removed all had similar mass surface densities which were close to the corresponding detection limit,  consistent with the suggestion that they were likely not real. The confirmed clouds ranged in mass between about 0.1 \msun\ and 2$\times$10$^7$ \msun\ and in radius between roughly 0.1 and 270 pc.
These ranges clearly include clouds smaller and less massive than GMCs  (i.e., R$_{GMC}$ $\gtrsim$ 3 pc, M$_{GMC} \gtrsim 10^3$ \msun) but they represent less than 8\% of the catalog clouds and have negligible influence on our analysis.

For comparison we also analyzed the GMC mass-size relation derived using dust extinction measurements obtained at optical and near-infrared wavelengths. For this purpose we examined two additional published data sets. First, we used a compilation of highly resolved (FWHP $\approx$ 2-3 arc min), 2MASS near-infrared extinction maps of 11 GMCs located within 0.5 kpc of the sun to define a benchmark Local GMC Sample (hereafter LGS; Lada et al. 2010, Lombardi et al. 2010, and references therein). These clouds were identified from wide-field extinction maps as contiguous regions with infrared extinctions (A$_K$) in excess of 0.1 magnitudes. The clouds in the LGS have well known distances and range in mass  from roughly 8$\times$10$^2$--10$^5$ \msun\ and in radii from about 3-20 pc. Second, we used the recent GMC catalog of Chen et al. (2020). This catalog contains 567 GMCs within $\sim$ 3 kpc of the sun; although a few are as distant as 3 kpc or more, the majority (80\%) are within 1.5 kpc of the sun. Thus, although the Chen et al. sample occupies a significantly larger volume of the Galactic disk than the LGS, it still represents a relatively local sample of clouds compared to those in the two CO catalogs. The Chen et al. clouds were identified from a dendrogram analysis of 3-D extinction maps that were presented in an earlier study (Chen et al. 2019); this study used optical data from Gaia DR2 (Gaia Collaboration et al. 2018) together with infrared data from 2MASS (Skrutskie et al. 2006) and WISE (Kirkpatrick et al. 2014) to construct fully sampled optical and infrared color-excess (extinction) maps of the entire Galactic plane. The extinction maps have an angular resolution of 6 arc min, comparable to the CO data. Distances to the extracted clouds were obtained using Gaia DR2 parallaxes and a modern variant of the Wolf (1923) method and ranged from roughly 0.36--3.6 kpc. The derived cloud masses ranged from 6 \msun\ to 8$\times$10$^5$ \msun\ with radii between 0.2--86 pc. 
As we did for the MD+17 catalog, we have retained in the Chen et al. catalog a small fraction of the clouds that are smaller and less massive than GMCs.

\section{Results and Analysis} 



\subsection{The \twco Mass-Size Relation for Milky Way GMCs}

\begin{figure}[t!]

\hskip -0.7in
\includegraphics[width=0.7\hsize]{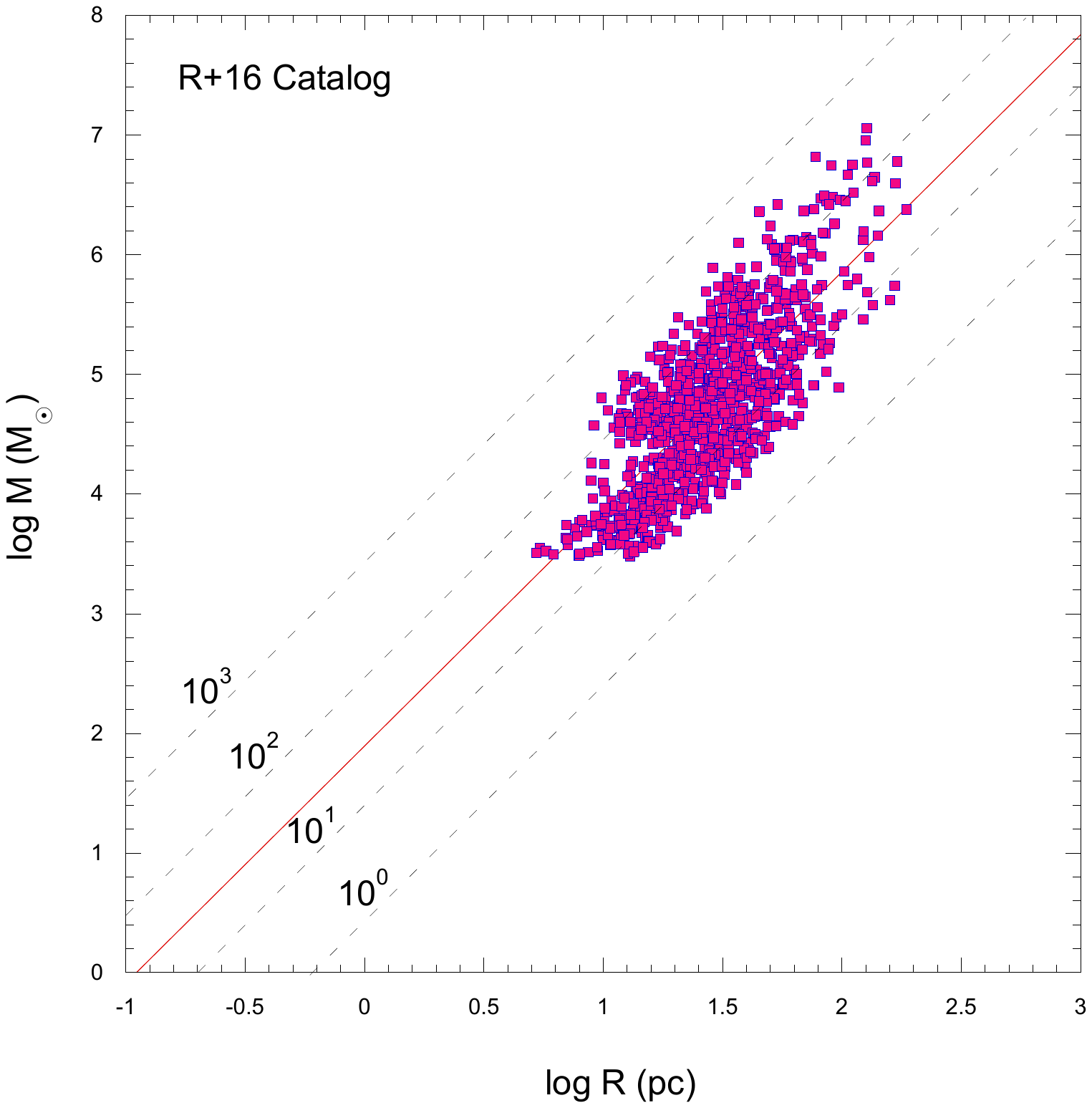}
\hskip -1.1in
\includegraphics[width=0.7\hsize]{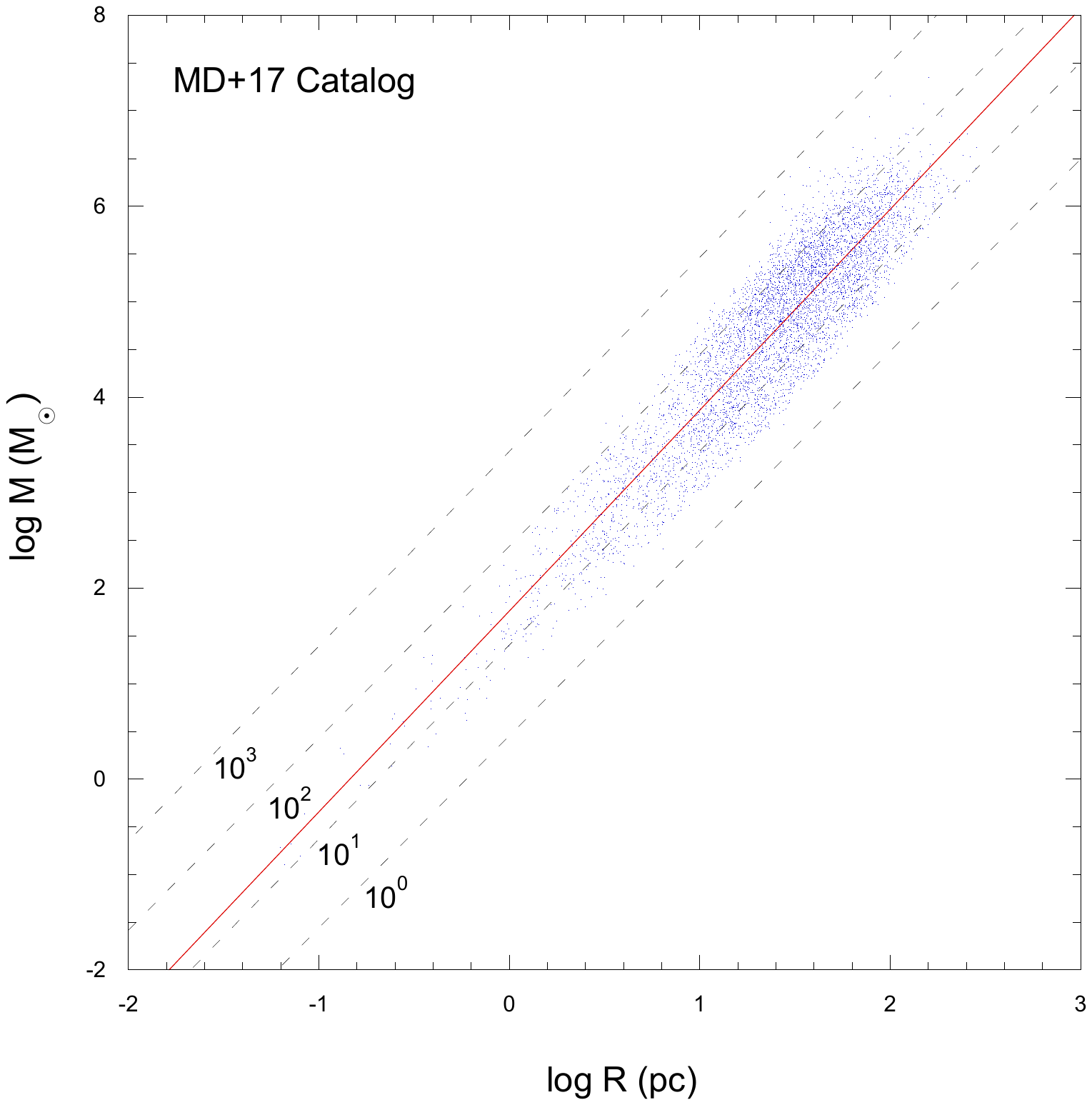}
\vskip -1.0 in
\caption{Mass-Size relations for Milky Way GMCs derived from $^{12}$CO observations.  The dashed lines in each plot correspond to locii of constant cloud surface densites (i.e., $\Sigma_{GMC}$ $=$ 1, 10, 100 and 1000 \msun\ pc$^{-2}$, respectively). The solid (red) lines are  linear least squares fits to the data.   The left panel plots GMCs from the cloud catalog of Rice et al. (2016). The right panel plots the GMCs from the cloud catalog of Miville-Deschenes et al. (2017). Both data compilations are characterized by distributions that are closely parallel to lines of constant surface density, clearly confirming Larson's (1981) relation, although with considerable scatter. Both catalogs are based on analysis of the data from the Dame et al. (2001) CO survey of the entire Milky Way.  \label{12MvsR}}
\end{figure}

In Figure \ref{12MvsR} we show the mass-size relations for the GMC populations identified in the R+16 and MD+17 catalogs.\footnote{Note that for all the data used in this study the cloud size is defined to be $R=\sqrt{Area/\pi}$ to facilitate comparisons.}   In both plots there is a strong correlation between the two quantities.  Moreover, the distribution of points are very nearly parallel to the lines of constant column density. A simple linear fit to the data finds 
${\rm log}M = 1.89 (\pm 0.07) + 1.98 (\pm 0.05) {\rm log}R$  or $M = 78.6  R^{1.98}$ for the  clouds in the R+16 catalog and 
${\rm log}M = 1.76 (\pm 0.02) + 2.10 (\pm 0.01)\ {\rm log}R$  
or $M = 57.5  R^{2.10}$ for the MD+17 clouds.  The fact that the power-law indices of both fits are very close to 2 nicely confirms Larson's scaling relation for Milky Way GMCs  implying that these clouds are indeed characterized by a constant surface density at least within the observed scatter. The values of these characteristic GMC surface densities can be approximately derived from the scaling coefficients, that is, $\Sigma_{GMC}$ $=$ 78.6/$\pi$ or 25.0 \msun\ pc$^{-2}$, and $\Sigma_{GMC}$ $=$ 57.5/$\pi$ or 18.3 \msun\ pc$^{-2}$, respectively and are in reasonable agreement with each other considering the difference in methods, etc.  However, the scatter is large with nearly identical logarithmic dispersion in both plots of $\approx$ 0.45 dex.  

The values of \sdsp derived from the scaling coefficients of the fits are smaller than the respective means of the individual data points, which are  $<$\sd$>$ $=$ 38.3 $\pm$ 43.4 and 40.7 $\pm$ 44.3 \sdu, where the quoted uncertainty is the standard deviation of the measurements in the two sets of data. This is somewhat surprising since  for a power-law scaling relation with a spectral index of $\sim$ 2 the two measures should be similar. However, the scaling coefficients of the fits are close to the median values of the two distributions, that is,  $\Sigma_{GMC}^{med} $=$\ $21.7  and 25.9 \sdusp for the Rice+10 and MD+17 catalogs, respectively. The higher values for the mean surface densities result from the presence of an extended high surface density tail in the observed frequency distribution of $\Sigma$s. This surface density tail consists of some of the most distant GMCs, located primarily in the inner regions of the Milky Way. Such measurements are biased upwards by the fact that for a fixed angular resolution and a fixed sensitivity there is a minimum cloud radius and mass that can be detected and measured at a given distance (e.g., Appendix C in Miville-Deschenes (2017)). Because of this, lower mass clouds are not detected at large distances and only the inner, high surface density, portions of the high mass clouds are observed. Thus, these measurements can overestimate the true mean surface density of the clouds and bias the calculation of the mean value.

\subsection{The Dust Mass-Size Relation for Milky Way GMCs}

\begin{figure}[t!]
\centering
\vskip -0.8in
\includegraphics[width=0.8\hsize]{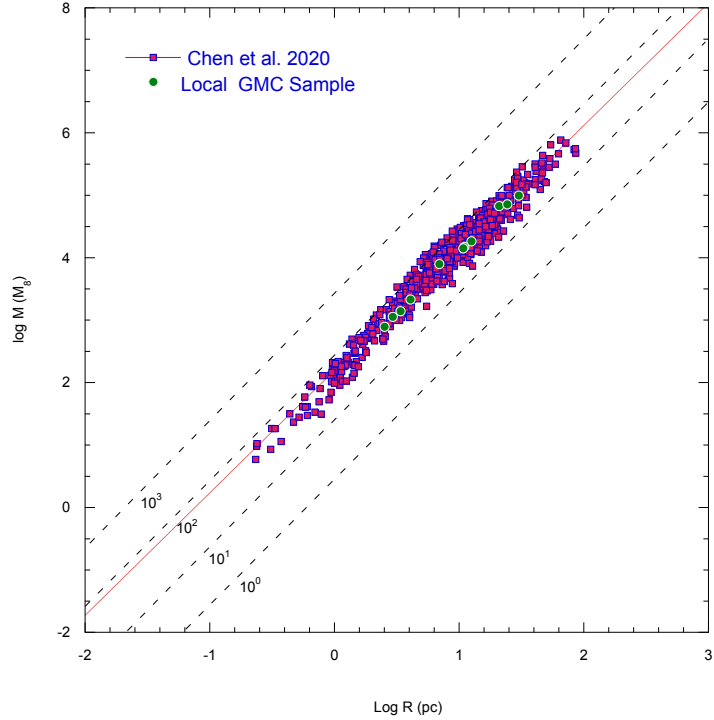}
\vskip -1.0 in
\caption{Mass-Size relations for Milky Way GMCs derived from dust extinction observations.  The dashed lines are as in Figure 1. The small (red) squares are clouds from the Chen et al. (2020) catalog while the (green) circles are clouds in the Local GMC Sample (Lada et al. 2010). The solid (red) line is a linear least squares fit to the Chen et al. data.  \label{DustMvsR}  }
\end{figure}

In figure \ref{DustMvsR} we show the GMC mass-size relation derived from extinction observations of the dust within the clouds. Filled square boxes are data from the Chen et al. (2020) survey and the filled (green) circles are the data from the LGS. The solid (red) line is a linear fit to the Chen et al. data which returns $ {\rm log}M = 2.19 (\pm 0.02) + 1.96 (\pm 0.02) {\rm log}R$  or $M = 155  R^{1.96}$ corresponding to a characteristic GMC surface density $\Sigma_{GMC} = $ 50 \msun\ pc$^{-2}$ for these clouds. A fit to the LGS data finds $ {\rm log}M = 2.12 (\pm 0.04) + 1.98 (\pm 0.04) {\rm log}R$  or $M = 131 R^{1.98}$ with $\Sigma_{GMC} =\ $42 \msun\ pc$^{-2}$. Unlike the CO results, these characteristic surface densities agree well with both the means of the individual GMC surface densities in the two samples (that is, $<$\sd$>$ $=$ 50.4 $\pm$ 20.9 and 41.2 $\pm$ 5.3 \sdu, respectively) and the median values ( $\Sigma_{GMC}^{med}$ $=$\  48.3 and 39.3) of the two extinction GMC samples.  Here again the quoted uncertainty is the standard deviation of the measurements. This agreement between the various estimates of \sd, is as expected for measurements without the distance bias mentioned above, a result of the fact that the two extinction samples occupy a smaller volume of the nearby Galactic disk.  

The different scalings exhibited by the two, extinction derived, mass-size relations ($\sim$ 50 \& 40 \sdu) are primarily a result of the different mass calibrations employed in the two studies. Specifically, Chen et al. (2020) use the calibration of Chen et al. (2015): N(H $+$ 2H$_2$) = 2.41 $\times$ 10$^{21}$  A$_V$ cm$^{-2} $, while the LGS adopts the more standard calibration: N(H $+$ 2H$_2$) = 1.87 $\times$ 10$^{21}$ A$_V$ cm$^{-2}$ (e.g., Lombardi et al. 2010; Bohlin et al. 1978). The ratio of the two calibration factors is 2.41/1.87 $=$ 1.29, and is comparable to the ratio of characteristic surface densities, i.e, \sd(Chen+)/\sd(LGS) = 1.25.  Once this is taken into account the two extinction studies are in excellent agreement with each other. The scatter in both dust data sets is significantly less than that observed in the CO relations. The scatter in the Chen et al. data is 0.18 dex and in the LGS only 0.04 dex, the latter an order of magnitude less than that found for the CO observations.  Moreover, the two dust studies produce measurements of the characteristic cloud surface densities (\sd) about a factor of 1.7-2 higher than indicated by the CO observations.

In this paper we will adopt the LGS gas-to-dust calibration. The reason for this is that the LGS extinctions are derived from ({\sl JHK}) infrared colors where the extinction law is not sensitive to variations R$_V$, the ratio of total to selective visual extinction (e.g, Mathis 1990). The Chen et al. extinctions, on the other hand, are primarily based on optical colors which are sensitive to variations in R$_V$ which are not uncommon.

\section{Nature of the Observed Scatter} 

\subsection{Experimental Uncertainties or Physical Variations?}

The significant difference in the size of the scatter between the \twco and extinction data is intriguing. It could be due to experimental uncertainties inherent in the \twco but not the extinction data and/or it could result from real, intrinsic variations in $\Sigma_{GMC}$ within the Galaxy that are not evident in the extinction data since those clouds occupy a smaller volume of the Milky Way.  Such Galactic variations would likely be systematic, perhaps environmentally driven, given the tight correspondence of the extinction data to a truly constant average GMC column density for local clouds (i.e., $<$\sd$>$ $=$ 41.2 $\pm$ 5.3 \sdu)\footnote{Unless otherwise stated, from here forward we will define the characteristic surface density, \sd,  to be equal to $<\Sigma_{GMC}>$, the mean of the individual surface densities in a sample under consideration. This would be the same value as ${1\over\pi} \times$ the coefficient, $a$, of a mass-size relation given by M = a R$^2$.}  However, the very large observed spread in the CO derived surface densities (i.e., 2 - 400 \sdu), if intrinsic,  would be difficult to reconcile with simple theoretical expectations of a constant surface density based on the observed structural properties of GMCs (Ballesteros-Paredes et al. 2012). 

Uncertainties in the distances to clouds are probably not a significant source of the error in derived values of  $\Sigma$, since $\Sigma$ $= M/R^2$ and both the numerator and denominator scale with D$^2$. Therefore, for mass-size relations that follow a $M \sim R^2$ scaling relation, distance uncertainties should not introduce any significant scatter away from the exact relation.  

One source of the large scatter could be related to cloud definition. In principle, CO clouds should have clear and definite boundaries prescribed by the photo-dissociation of molecules due to UV radiation at the cloud edges. One expects the (CO) molecules to be  almost fully dissociated at extinctions of A$_V$ $\approx$ 0.5 magnitudes corresponding to $\Sigma$ $\approx$ 10 \sdu (van Disheock and Black 1988; Visser et al. 2009; Wolfire et al. 2010). This appears to be the case for the local cloud sample where individual clouds are relatively isolated from one another on the sky. However, for more distant clouds and most of the clouds inside the solar circle we expect to observe  frequent and often significant overlap of clouds along the line-of-sight. How to separate and extract cloud properties in such circumstances is a vexing problem that has hampered the field for decades.
One advantage offered by spectroscopic observations is that overlap on the sky can be potentially resolved in velocity space, thus spurring the development of a number of automated, three dimensional, cloud identifying algorithms such as, CLUMPFIND (Williams et al. 1994)  and CPROPS (Rosolowsky and Leroy 2006) and techniques such as dendrogram analysis (Rosolowsky et al. 2008) and gaussian decomposition with heirarchical cluster identification (MD+17).  These methods are complex and it is difficult to compare their relative efficacy since few have been applied to the same data or calibrated by application to an agreed upon standard. Furthermore, it is exceedingly difficult to cross correlate the properties of individual clouds between the various catalogs.

It is of interest, from this standpoint, to compare the R+16 results to those of  MD+17. These two studies applied different methodologies to identify and extract GMCs from the same CO survey. As mentioned above the R+16 study used a dendrogram analysis while MD+17 employed a gaussian decomposition based hierarchical cluster finding approach to extract clouds from the DHT survey. These two approaches yielded very different results for numbers of GMCs identified  (1037 vs 5577), the range in masses (10$^3$ - 10$^7$ vs 10$^{-1}$ - 10$^7$ \msun) and sizes (2-240 pc vs 0.1-270 pc) of the extracted clouds as well as the fractions of the DHT survey recovered (40\% vs 98\%) Yet,  despite these differences, analysis of both data sets produce very similar results for the slopes, scalings and scatter of the respective mass-size relations derived from the data. Evidently, for the \twco data, the measured scatter in the mass-size relation does not appear to be particularly sensitive to differences in the methodologies used to identify the clouds in these two studies. This suggests that other systematic effects dominate the scatter for this tracer.

\begin{figure}[t!]
\centering
\vskip -0.8in
\includegraphics[width=0.75\hsize]{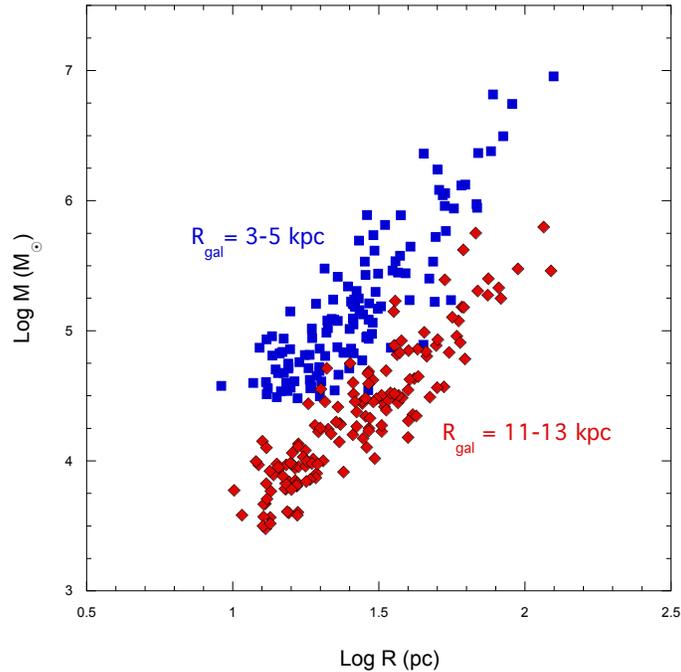}
\vskip -1.0 in
\caption{Mass-Size relations for Milky Way GMCs in the inner vs outer regions of the Galaxy taken from the GMC catalog of Rice et al. (2016).  The (red) diamonds correspond to clouds in the outer Milky Way, located between 11-13 kpc from the center of the Galaxy. The (blue) squares correspond to clouds in the inner Milky Way, between 3-5 kpc from the center of the Galaxy.  \label{ioMvsR} }
\end{figure}

\subsection{A Radial Variation Across the Galaxy}

Another possible explanation for the differences between the \twco and dust derived mass-size scaling relations may be tied to the fact that the \twco surveys cover a much larger volume of the Milky Way than the dust extinction observations. To test this hypothesis we examined the dependence of the mass-size relation on Galactocentric radius. In Figure \ref{ioMvsR} we plot the mass-size relation for sub populations of GMCs from the R+16 catalog located in the inner and outer Galaxy. Both sub populations clearly display independent correlations between their masses and sizes with only slightly differing slopes  (2.18 and 1.95 for the inner and outer GMCs, respectively). However, there is a significant difference in the scaling between the two mass-size relations, corresponding to differing values of $\Sigma_{GMC}$. Indeed, their mean surface densities are quite different being 84.5 \sdu\  and 11.7 \sdu\  for the inner and outer GMCs, respectively. 

 The scatter (0.22 dex) in the inner galaxy surface densities is comparable to that (0.28 dex) of the outer Galaxy GMCs. But both are significantly less than that derived for the whole MW sample (0.43 dex) and closer to that (0.18 dex) of the Chen et al. extinction sample. The same analysis of the MD+17 sample returns very similar results. {\it This indicates that a significant amount of the observed scatter in the Galaxy-wide  \twco mass-size relation is due to systematic variations in \sd\ with Galactic radius.} These facts suggests that a varying scaling  parameter  may be needed to describe the observed mass-size relation for Milky Way GMCs.
It is interesting to note here that the mean surface density of the outer Galaxy clouds \sd\ $\sim$ 12 \sdu\ corresponds to a visual extinction of only $\sim$ 0.5 magnitudes, very close to the value for molecular dissociation in the MW suggesting that these clouds are very tenuous objects. Although, as will be discussed later, this could also indicate issues with the adopted \twco mass calibration. Indeed, this is suggested by the fact that  when considering the whole cloud sample there still remains a significant difference in the scaling factor (i.e., \sd) between the CO and dust relations with the CO derived \sd s being a factor of 2 or more lower than those derived for the dust.

\begin{figure}[t!]
\centering
\vskip -0.8in
\includegraphics[width=0.9\hsize]{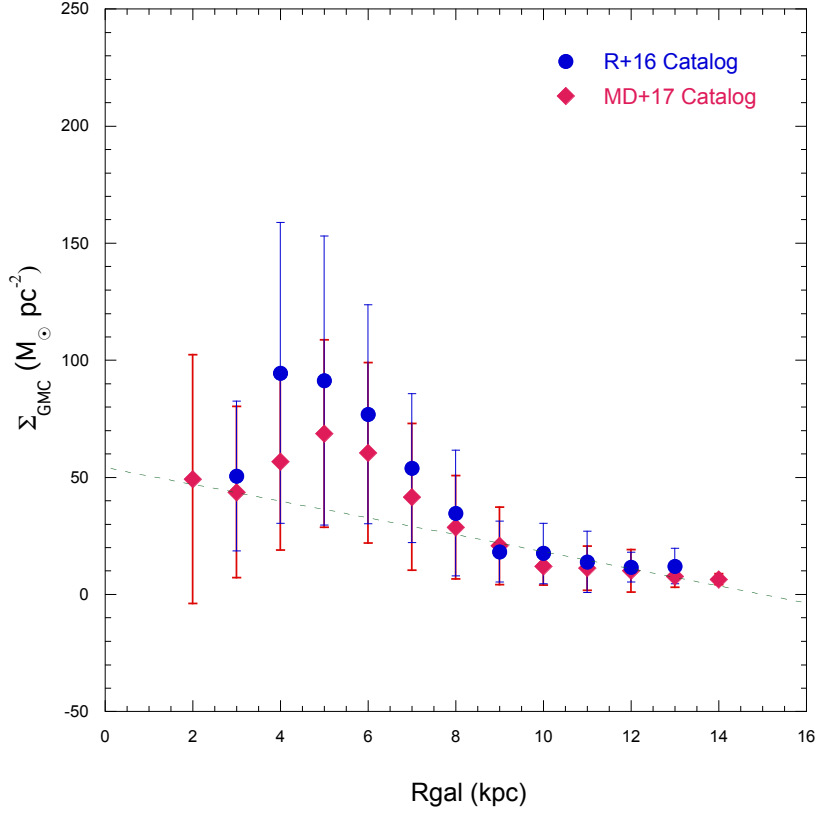}
\vskip -1.0 in
\caption{The radial dependence of mean GMC surface densities across the MW disk. The (blue) circles are data from the Rice et al. (2016) GMC catalog and the (red) diamonds are based on data from the Miville-Deschanes et al. (2017) catalog. The error bars represent the standard deviation of the values. The light dashed line is a least squares fit to the points excluding those between 4-8 kpc. (see text)  \label{SDvsRgal} }
\end{figure}

To further investigate the variation of \sd\ with location in the Milky Way we calculated the mean of the GMC gas surface densities  as a function of Galactic radius for both \twco catalogs. The results are plotted in Figure \ref{SDvsRgal}. To avoid confusion we emphasize here that our graph plots the mean of the gas surface densities of individual GMCs and not the azimuthally averaged surface density of molecular gas as a function of Galactocentric radius, though the two distributions are undoubtedly related.
In the outer Galaxy the data from the two catalogs  agree well and the relation is relatively flat from 9 to 14 kpc. The GMCs in the outer Galaxy appear to be characterized by a constant \sd\ $\approx$ 13 $\pm$ 4 \sdu. Inwards of 9 kpc both relations begin to rise as the radius decreases, until around 4 kpc where the points decline with decreasing radius. The broad peak in the distributions between 4-7 kpc corresponds to the well-known molecular ring which contains the inner Scutum-Centaurus and Norma spiral arms. The plot shows clear evidence for a systematic variation in \sd\ with Galactic radius. 

\subsubsection{Nature of the Radial Variation: A Baseline Linear Gradient}

The question we now consider is whether the observed variation with Galactic radius is intrinsic  or produced by an unrecognized systematic error in the CO observations. That GMCs have power-law pdfs almost guarantees that they should have the same column density if measured from the same boundary threshold. This is because the PDFs of GMCs peak at low surface densities, near those of the boundary thresholds, and then fall off very steeply to higher column densities that occupy significantly smaller areas of the cloud. It would require significant internal changes in cloud structure for \sdsp not to be constant (Ballesteros-Paredes et al. 2012). However, changes in the measurement threshold result directly in changes to the  individually measured values of \sd. As mentioned earlier, Lombardi et al. (2010) showed that for the clouds in the LGS,  \sd\ $\approx$ 2 $\times \Sigma_{threshold}.$  GMCs in the LGS are well separated on the sky and have relatively well defined boundaries. The same is mostly true for clouds observed from our vantage point in the outer galaxy. 

However, in the molecular ring region the space density of clouds is much higher, resulting in significant blending of cloud emission both on the plane of the sky and in velocity.  The velocity blending is caused not only by the well-known distance ambiguity in the inner Galaxy, but also by the significant internal velocity dispersion of GMCs, their random and non-circular streaming motions, all coupled with a shallow gradient of velocity with distance near the tangent points in the inner Galaxy.   At a typical direction in the inner galaxy, we find that a cloud at the tangent point and another 1 kpc closer along the line-of-sight will differ in velocity by only $\sim$ 5 km s$^{-1}$ owing to Galactic rotation (using the universal rotation curve of Reid et al. 2014). This is comparable to the internal velocity dispersion of a 10$^5$ \msun \ GMC (Dame et al. 1986; Solomon et al. 1987), typical of the clouds studied here (e.g., Figure 1), and will result in heavy overlap along the line-of-sight. 

In this situation, finding consistent outer boundaries of the GMCs is challenging for any cloud identification algorithm. This can result in significant variations in the level of the measured boundaries and introduce a bias toward higher values of the measured \sd\ in these directions.  In addition, as mentioned earlier, sensitivity and angular resolution limits can also bias GMC measurements to higher values for the more distant clouds in these directions. This effect is difficult to quantify and we will not attempt to do so here. However,  Ballesteros-Paredes et al. (2019) modeled the effect of overlapping GMCs on the mass-size relation in CO observations and found that random variations in the degree of cloud overlap in a large cloud sample could, in addition to artificially  increasing the slope of the mass-size relation, also produce much of the scatter seen in the relations. 
 
Inward of $\sim$ 4 kpc the overall space density of clouds drops significantly probably owing to the presence of the Galactic bar (Blitz \&Spergel 1991). In this region the clouds are again fairly well defined as they are at and beyond the solar circle.
We take advantage of this situation to produce a conservative measure of the variation of \sd\ with Galactic radius. For this exercise we consider only the MD+17 data since it spans a larger range of Galactic radius. We perform a least squares fit to the MD+17 data, excluding the points between 4-8 kpc to estimate the minimum (unblended) radial gradient in \sd. We find that: 

\begin{equation}
 \Sigma_{GMC}(R_{gal}) = 54.5 - 3.7 R_{gal} \ \    {\rm M_\odot pc^{-2}}. 
 \end{equation}
 
 \noindent
 where $R_{gal}$ is the radial distance (in kpc) from the center of the Galaxy.
 The fit is shown as a dashed line in Figure \ref{SDvsRgal}. Since the relation in equation 1 was derived using data least likely to be biased by crowded and overlapping clouds, we suggest that it represents a baseline radial gradient in unblended GMC surface densities. 
\subsubsection{Nature of the Radial Variation: A Metallicity Dependent CO X-factor}

However the assumption of a constant CO mass calibration factor (X$_{CO}$) that was used in both catalogs, could introduce a systematic error in the derived \sd. In particular, this assumption could play a role in producing the baseline radial variation in \sd\  shown in Figure \ref{SDvsRgal} and described by equation 1. This is because the CO abundance and consequently  X$_{CO}$ are believed to be functions of metallicity (Bolatto et al 2013) and the Milky Way is known to have a radial metallicity gradient  (Maciel \& Costa 2010).  The Galaxy's metallicity, 
decreases with Galactic radius out to about 10 kpc, where it appears to flatten (Maciel \& Costa 2010 and references therein).  This is very similar to the behavior of \sd(R$_{gal}$) seen in Figure \ref{SDvsRgal}, suggesting that \sd(R$_{\rm{gal}})$ varies with metallicity and that the gradient we measure in \sd(R$_{\rm{gal}}$) could  reflect the use of a constant value of \xco in the surface density calculations rather than a value adjusted for metallicity. 
Indeed, studies indicate that the metallicity gradient in the Milky Way can be expressed as log(Z/Z$_\odot$) $\sim$ -0.06 (R$_{\rm{gal}}$) (Genovali et al. 2014; Wenger et al. 2019) giving ${Z(3kpc)\over{Z(10kpc)}}$ = 2.51 which is essentially the same as the ratio of mean GMC surface densities given by equation 1, ${\Sigma(3kpc)\over{\Sigma(10kpc)}} = $ 2.48.   
Apparently the baseline gradient in \sdsp that we observe in Figure \ref{SDvsRgal} could be largely accounted for by correcting the values of the masses and surface densities  by a metallicity dependent conversion factor, that approximately behaves as X$_{CO}$(Z) $\sim$ Z$^{-1}$. Such a metallicity dependence of  \xco is consistent with predictions of a recent set of simulations by Feldmann et al. (2012). Because of this dependence, the presence of the baseline gradient in \sdsp by itself does not provide convincing evidence for the need of a  variable scaling parameter in the mass size relation, at least for those clouds outside the molecular ring.

\begin{figure}[t]
\centering
\vskip -0.8in
\includegraphics[width=\hsize]{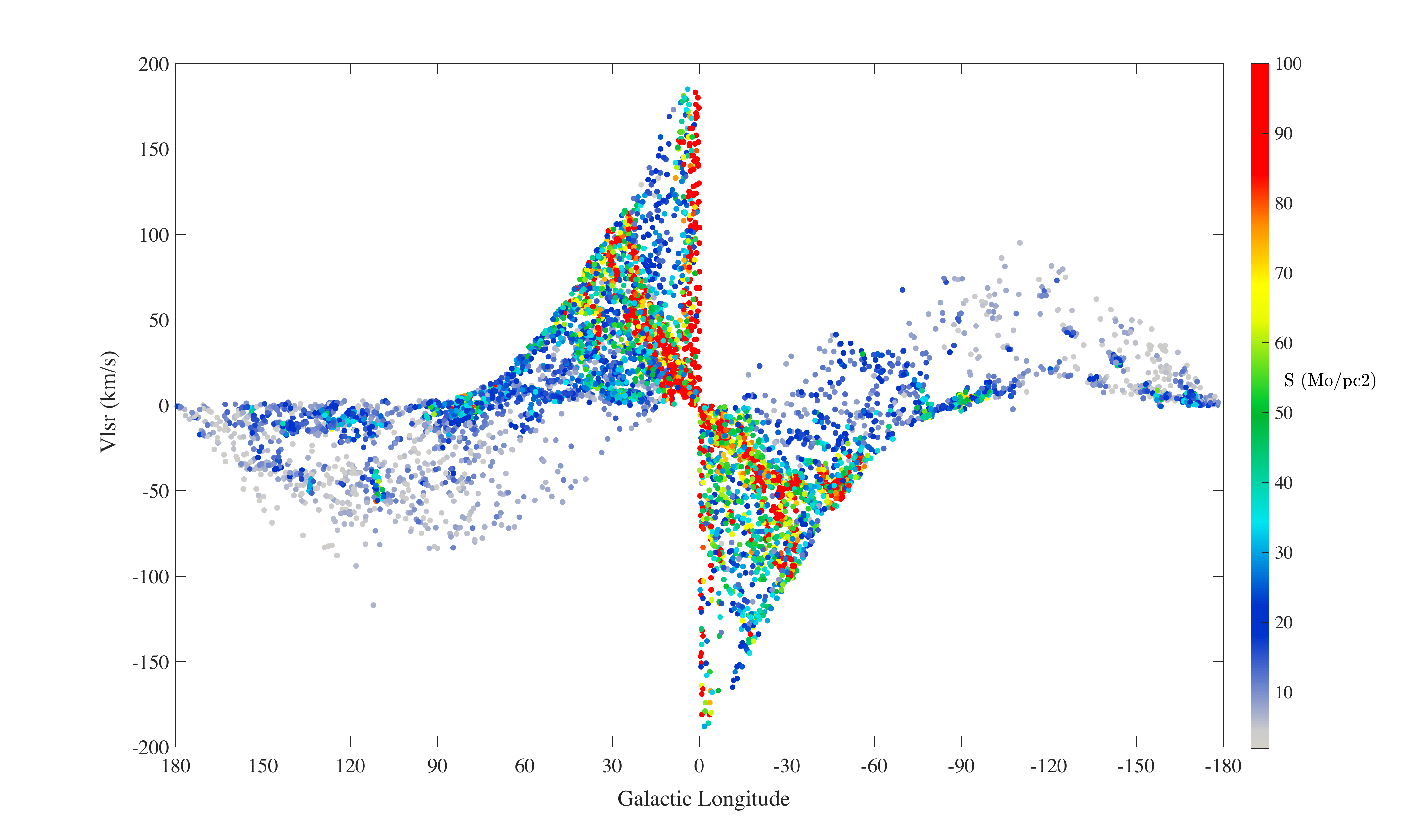}
\caption{The distribution of GMCs from the Miville-Deschanes et al. (2017) catalog in velocity-galactic longitude space showing their surface densities as an added (color) dimension. The color bar illustrates the dynamic range of \sd. The highest surface density (red-yellow) clouds are seen to clearly trace out a locus in the inner Galaxy that is expected for spiral arms. This is the region of the inner Scutum-Centaurus and Norma arms, and is coincident with the area known as the molecular ring. (see text)  \label{lvmap} }
\end{figure}

\subsubsection{Nature of the Radial Variation: A Peak in the Molecular Ring}

The broad peak between 4 -- 8 kpc in Figure \ref{SDvsRgal} does suggest a significant departure from an otherwise constant GMC surface density across the Milky Way. To better understand the nature of this peak we plot the positions of the GMCs in the DM+19 catalog on the longitude-velocity diagram in Figure \ref{lvmap}. The clouds with the highest surface densities (red and yellow symbols) are found to lie in two regions, the first a locus between $l= \pm$ 30$\degr$, similar to that expected for an inner spiral arm or arms and the second a vertical distribution near $l =$ 0.0$\degr$.  The surface densities measured in this latter region are almost certainly spurious, since the X-factor in the Galactic center region is thought to be up to 10 times lower than elsewhere in the disk (Bolattto et al. 2013). The Scutum-Centaurus and Norma arms lie in the former region which makes up the molecular ring. This suggests perhaps that  the surface densities of the GMCs are enhanced by some process in the spiral arm structures.
If so, this would indicate again the need for a variable (environmentally dependent) scaling parameter for the mass-size relation.  
However, as mentioned earlier, in this direction on the sky we expect to observe crowding and overlap of the clouds along the line-of-sight where biases in the measured surface density can be introduced by cloud extraction algorithms.

How much of the observed enhancement in the molecular ring is due to a change in the physical properties of the GMCs within the inner spiral arms and how much is a result of a bias in the measured surface densities due to significant cloud overlap and blending? The following considerations suggest that the blending of clouds could produce both increased scatter and upward bias in \sdsp for GMCs in the molecular ring.
First, when emission from a cloud is superposed on a varying background due to crowding and blending the identification algorithms do not always measure cloud properties to the true (or same) cloud edge or boundary. This produces an upward bias in the measured surface densities  because the individual GMCs are generally stratified with power-law pdfs such that the inner regions are characterized by much higher surface densities and smaller total surface area than the outermost regions.
Second, this bias would be expected to increase with distance through the molecular ring  as decreasing spatial resolution increases cloud blending. This would produce an artificial increase in the measured surface densities with decreasing Galactic radius in the ring region.  Third, the average scatter in the calculated mean surface density in the molecular ring GMCs  (42 \sdu) is found to be 4.5 times higher than that (9 \sdu) in the outer Galaxy GMCs. Random variations in the degree of cloud overlap could account for the larger dispersions measured for \sdsp in the molecular ring compared to other regions of the Galaxy (e.g., Ballesteros-Paredes et al. 2019). These considerations add to the surmise that surface density measurements made toward the molecular ring are much more susceptible to systematic biases that both increase the scatter in and the measured values of \sdsp  compared to other regions of the Galaxy.  

Unfortunately, it is difficult to more precisely determine the magnitude of these effects from the published CO studies. Though the rise in measured GMC surface densities in the molecular ring suggests a departure from constant surface density clouds and the Larson relation in that region, the magnitude of this departure is unclear at the present time and it could be much smaller than current measurements imply.

\begin{figure}[t!]
\centering
\vskip -0.8in
\includegraphics[width=0.8\hsize]{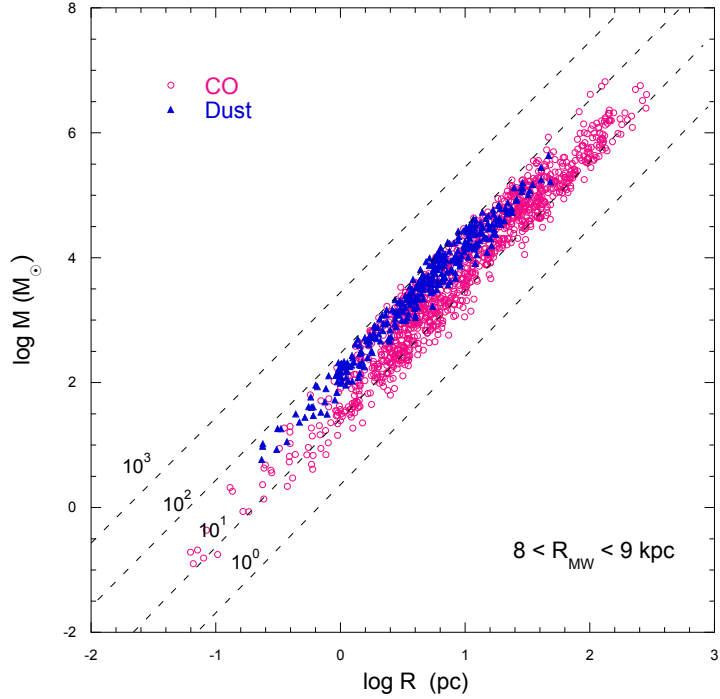}
\vskip -1.0 in
\caption{Comparison of the mass-size relations derived from CO and dust observations for GMCs between galactic radii of 8-9 kpc. The open circles represent CO data from the Miville-Deschanes et al. (2017) catalog while the solid triangles represent extinction data from the Chen et al. (2020) catalog. The dashed lines are loci of constant surface density from 10 to 1000 \sdu.   \label{DGMvsR} }
\end{figure}

\subsection{Comparison of Gas and Dust at 8.5 kpc:  Calibration of the CO X-factor}

Given the radial dependence of the CO-derived GMC surface densities, it is of interest to investigate 
 the mass-size relations for a sample of clouds all at the same Galactic radius. In figure \ref{DGMvsR} we plot the mass-size relations for GMCs between 8-9 kpc from the Galactic center derived from both CO (MD+17) and dust extinction (Chen et al. 2020). It is clear that the dust observations are characterized by considerably less scatter than the CO observations. Moreover, the characteristic GMC surface density, \sd, or overall scaling for the two relations is clearly different for the two data sets. The mean surface densities are 51.5 $\pm$ 1.1 and 26.6 $\pm$ 0.6 \sdu, for the extinction and CO-derived measurements, respectively. Here the quoted uncertainties are the standard errors in the mean values. This indicates that there is a systematic difference in the total (H$_2$)  column density calibrations of a factor of $\sim$ 1.9. Additionally we tired further restricting this group of clouds to those within 3 kpc  of the sun
 to better match the extinction catalog. The resulting mean surface density for the CO clouds (\sdsp $=$ 26.8 $\pm$ 0.7) was essentially identical to that of the more expanded sample. As mentioned earlier the mass-size relation for the Chen et al. dust clouds is in excellent agreement with the mass-size relation of the LGS once differences in assumed gas-to-dust ratios are taken into account. In this paper we adopt the calibration of LGS as the fiducial calibration.  For the LGS calibration  \sd = 41.2 $\pm$ 1.6 \sdu. Here the quoted uncertainty is the standard error in the mean. This implies a calibration difference between the GMCs in the \md CO catalog and the LGS extinction derived dust column density measurements of a factor of 1.55 $\pm$ 0.07. Similarly for the GMCs in the \rice catalog between 8 - 9 kpc from the Galactic center (where \sd $=$ 20.3 $\pm$ 1.3 \sdu), we find a calibration difference of a factor of 2.03 $\pm$ 0.15 with respect to the LGS measurements.

The difference between the CO and the LGS dust mass calibrations  provides a direct  measure of the CO conversion factor (\xco) for the Milky Way GMCs between 8 $\leq R_{gal} < $9 kpc.  Both \twco catalogs adopted the standard X-factor for their mass calibration (i.e.,  \xco = 2 $\times$ 10$^{20}$ cm$^{-2}$ (K-km/s)$^{-1}$;  Bolatto et al. 2013) and comparison with the extinction data indicates that this underestimates the masses by a factor of 1.55-2.03. Correcting for this deficit using the average of the DM+17 and R+16 data we derive a CO mass calibration factor of: $$ {\rm X_{CO}(R_\odot)} = 3.6 \pm 0.3  \times 10^{20} \ \  {\rm cm}^{-2} {\rm (K-km/s)}^{-1}$$
\noindent
We emphasize that this calibration applies near the solar circle (i.e., R$_\odot$) and is  appropriate for the two CO catalogs considered here. It may not necessarily apply to CO observations made using different data obtained on different (particularly subcloud) spatial scales and/or analyzed using different methodologies for cloud identification and extraction (e.g., Lombardi et al. 2007, Pineda (2008), Kong et al. 2015, Lee et al. 2018, Lewis et al. 2020).

Earlier we suggested that the underlying radial gradient in cloud surface densities derived from CO (i.e., Eq.1 and dashed line in Figure 4) might wholly be the result of an unaccounted for variation of \xco with metallicity.  If this is the case  then \xco must vary as the inverse of the mean cloud surface densities:
$$X_{CO}(R_{gal}) =\ X_{CO}(R_\odot){\Sigma(R_\odot) \over{ \Sigma(R_{gal}).}}$$ 
\noindent
The value of X$_{CO}(R_\odot)$ derived above together with Eq 1. for $\Sigma(R_{gal})$ yield the following expression for \xco as a function of Galactic radius:
\begin{equation} 
 \rm{X_{CO}(R_{gal}) = \left[{83\over{(54.5 - 3.7 {\rm R_{gal}})}}\right]  \ for \ 2 < R_{gal} \leq10 \  kpc}  \ \ 
\end{equation}

\noindent
where  \rgal \ is in units of kpc and \xco in units of 10$^{20}$ cm$^{-2}$ K$^{-1}$ km$^{-1}$ s.
For R$_{\rm gal}$ $>$ 10 kpc  we assume a constant \xco =~6.0 to reflect a constant or more slowly varying metallicity in the outer Galaxy.

We emphasize here that our determination of \xco(R$_{\rm{gal}}$) is an approximate one. It is based on our inference that \xco \ inversely varies with metallicity and thus with \sdsp in order to account for the gradient in Equation 1. The facts that
${\Sigma(3kpc)\over{\Sigma(10kpc)}}$  $\approx$ ${Z(3kpc)\over{Z(10kpc)}}$  and that the inferred metallicity dependence of \xco  is consistent with the predictions of recent simulations (Feldmann et al. 2012), provide intriguing evidence in support of the idea that a radially varying, metallicity dependent, X-factor might solely explain the surface density gradient of unblended clouds given by Equation 1. 

\section{Discussion}

Our analysis shows that catalogs derived from  two independent tracers of molecular gas,  \twco and dust extinction, yield mass-size scaling relations with M$_{GMC}$ $\sim$ R$_{GMC}^{2}$ which implies constant average surface densities for the corresponding cloud populations. However, the scaling coefficients of the CO and extinction derived relations significantly differ, implying  different mean GMC surface densities for the two tracers. Moreover, the scatter characterizing the CO relations is significantly larger than that of the  extinction relations. We find that the increased scatter of the CO mass-size scaling relations is a result of a radial variation in the scaling coefficient of the relations or equivalently in the characteristic "constant" GMC surface density implied by them. This radial variation is not apparent in the extinction observations because they do not span a sufficiently large range in Galactic radius. Calibration of the CO surface densities using the dust extinction measurements of local GMCs results in a larger X-factor than that used in the published CO catalogs and accounts for the different scaling coefficients found for the published CO and extinction catalogs.  

\begin{figure}[t!]
\centering
\vskip -0.8in
\includegraphics[width=0.9\hsize]{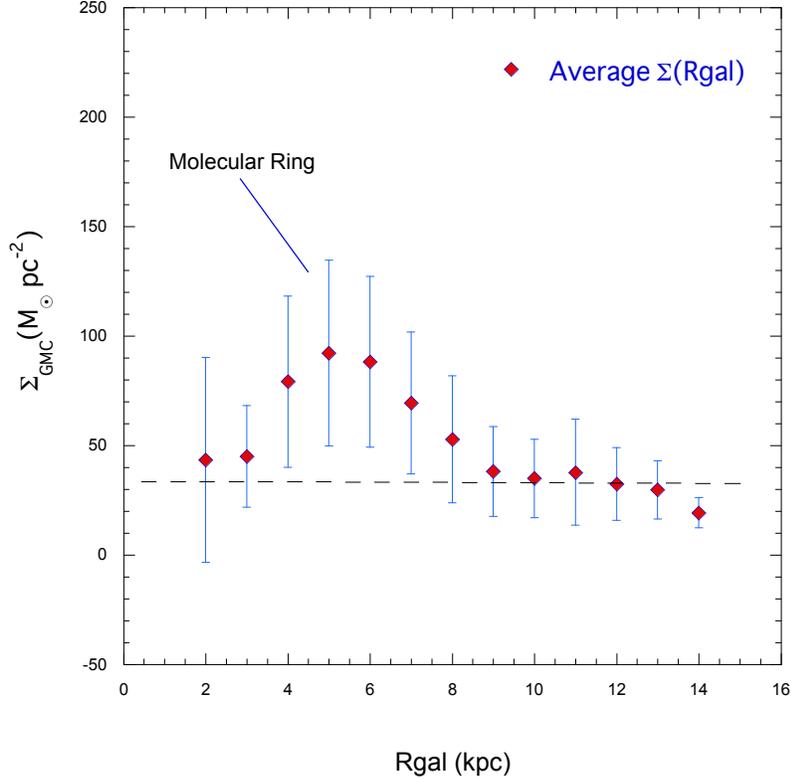}
\vskip -1.0 in
\caption{The radial dependence of mean GMC surface densities across the MW disk corrected for a metallicity and radially dependent X-factor. The data represent averages of data from the two CO catalogs considered here.  The error bars represent the standard deviation of the values and in the outer Galaxy are dominated by the systematic differences in derived surface densities between the two CO surveys. The light dashed line is the mean surface density derived for clouds outside the molecular ring (see text).   \label{SDvsRgalcorr} }
\end{figure}

We decomposed the radial variation of \sdsp into two components: a linear gradient of decreasing \sdsp with Galactic radius and a broad peak that coincides with the molecular ring in the inner Galaxy and is superposed on the linear gradient. We noticed that the linear decrease with radius is similar to that for metallicity in the Milky Way and posited that it results from the use of a constant, rather than metallicity dependent,  X-factor for computing masses from CO. In equation 2 we derived an approximate form for the variation of \xco with \rgal. 
To provide our best estimate for the radial variation of \sdsp in the Galaxy we correct the mean CO derived surface densities by the radially dependent X-factor derived in equation 2 and plot the result in Figure \ref{SDvsRgalcorr}.  The individual data points are derived from averages of the R+16 and \md catalogs. The error bars include both the intrinsic dispersions in the individual data sets and the systematic differences in the derived surface densities between the two catalogs but do not include any systematic error in X$_{CO}$. 

The figure shows that a single mass surface density of \sd $=$ 35 \sdusp well describes the clouds outside the molecular ring. The surface densities of clouds in the molecular ring appear to significantly deviate from this otherwise Galaxy-wide constant value.
The average surface density within the molecular ring is found to be $<$\sd$>$ $=$ 82 $\pm$ 10 \sdu, a factor of  2.3 higher than that for clouds outside the ring ($<$\sd$>$ $=$ 35  $\pm$ 8 \sdu).  As discussed earlier this difference is likely an upper limit because of  a significant upward bias in the measured surface densities in the molecular ring due to cloud blending. 

Although the size of the peak in GMC surface density in the molecular ring  is very uncertain, it is possible that environmental conditions in this region account for at least some of the increase in the mean surface densities there. 
In particular, comparison of GMC mass-size relations in nearby disk galaxies suggests that the scaling coefficient of the mass-size relation can vary between galaxies and that the mid-plane pressure in a galactic disk is directly related to its \sdsp (Faesi et al. 2018). This could occur if the pressure within a GMC was in equilibrium with the external mid-plane pressure of a galactic disk   as will be discussed in more detail below. The weight of a self-gravitating cloud gives rise to an internal pressure that only depends on its surface density, P$_{GMC}$ $\propto$ G$\Sigma_{GMC}^2$ (Bertoldi \& McKee 1992). The ratio of internal pressures between GMCs in the ring and the outer galaxy is then  $\Sigma_{ring}^2$/$\Sigma_{outer}^2$. 
For the surface density ratio for molecular ring and outer Galaxy GMCs calculated above, the corresponding ratio of internal pressures would be $\Sigma_{ring}^2$/$\Sigma_{outer}^2$ $=$ 5.5 $\pm$ 1.4.  This value is likely an upper limit to the true value, given the bias embedded in the measurements of \sdsp in the molecular ring discussed earlier. 

To estimate the external  pressure of the Milky Way disk acting on the clouds we assume that this pressure, $P_{ISM}$, originates from two components,  the stellar potential of the disk and the weight of the atomic (HI), interstellar gas. We follow the analysis of Faesi et al. (2017) and write the mid-plane pressure as: $$ P_{ISM} = {{\pi G}\over {2}}  \Sigma_{HI}\left[ \Sigma_{HI} + \sigma_{HI}{\Sigma_* \over{\sqrt{2} \pi G h_*}}\right]$$
\noindent
where $\Sigma_*$ is the stellar surface density, $\sigma_{HI}$ is the HI velocity dispersion, and $h_*$ is the stellar scale height (see also Elmegreen 1989 and Blitz and Rosolowsky 2004). We assume h$_*$ $=$ 300 pc (Momany et al. 2006) and that $\Sigma_{HI}$ = 11.0 \sdusp (McKee et al. 2015; Kalberla \& Dedes 2008) and $\sigma_{HI}$ = 9 km s$^{-1}$ (Malhotra 1995) and both are constant between 4-13 kpc. 
We determine $\Sigma_*$ assuming a value of 33 \sdusp at the solar circle (McKee et al. 2015) and the exponential stellar mass surface density profile of Kent et al. (1991). 
We compute the ratio of $P_{ISM}$ between R$_{\rm{gal}}$ $=$ 5 and 11 kpc to be ${P(5 kpc)}\over{P(11 kpc)}$ $=$ 1.9, a factor of $\sim$ 2.9 lower than needed to explain the increased surface densities in the molecular ring compared to those in the outer Galaxy.  An independent estimate of the radial pressure profile of the Milky Way disk calculated by  Wolfire et al. (2003) results in an expected ratio of pressures, ${P(5 kpc)}\over{P(11 kpc)}$ $=$ 3.0, somewhat higher than our estimate but still below that required to explain the increased surface densities in the molecular ring. Moreover, the fact that the GMC surface densities drop inward of 4 kpc also suggests that an inwardly increasing gradient in the mid-plane pressure is unlikely to be solely responsible for the increased surface densities in the molecular ring.
If the true \sdsp in the molecular ring is close to the upper limit we estimated here, then another source of pressure (e.g., spiral arm shocks) within the molecular ring would be needed to explain the observations. It is also possible that the clouds in the molecular ring are over pressured with respect to the external mid-plane pressure because they are strongly self-gravitating and perhaps even critically unstable.  We hesitate here to speculate any further on the cause of the increased surface densities of GMCs in the molecular ring since the actual magnitude of the increase is so uncertain at the present time. Additional modeling along the lines of the Ballesteros-Paredes et al. (2019) study might shed more light on this issue.

Finally, we note that although  the radial variation in GMC  surface densities can account for about half of the scatter in the Galaxy-wide mass size relations derived from CO observations (i.e., Figure 1) the remaining scatter is still significant.  
One source of the scatter could be the inability of the cloud extraction and identification methods to measure consistent cloud boundaries or measurement thresholds.  
As we discussed earlier, the derived average surface density of a cloud scales with the choice of the outer boundary surface density threshold (Lombardi et al. 2010). 
Measurement of outer boundary thresholds are only available for one of the studies considered here. Chen (2020; personal communication) has provided us with the thresholds used in the dendrogram decomposition of the Chen et al. extinction survey. These thresholds span a dynamic range of about a factor of 5 in extinction. The dispersion in the logs of these threshold boundaries is found to be 0.23 dex, comparable to the level of scatter (0.18 dex) in the corresponding mass-size relation. Thus, at least for this catalog, the measured scatter in the relation could arise from the variation in the adopted cloud boundary surface densities. Similar effects are likely in the CO data. In particular, the R+16 catalog also used a dendrogram cloud extraction methodology. The LGS measurements  employed the same boundary level (A$_K$ = 0.1 mag) for all the clouds and the small scatter in the corresponding measurements of \sdsp (0.04 dex) is likely due in part to this fact.
The modest difference in scatter between the radially resolved CO and extinction relations (0.25 dex vs. 0.18 dex) could be due to a number of factors. One possibility is intrinsic cloud-to-cloud variations in the X-factor. Such variations might arise due to environmental variation in cloud scale properties such as temperature structure and depletion (e.g., Kong et al. 2015, Lewis et al. 2020). 

\section{\label{conclusion} Summary and Conclusions}

We have analyzed molecular cloud catalogs from the literature derived from both \twco and dust extinction observations to test Larson's (1981) original finding that $M_{GMC} \sim R_{GMC}^2$ and the consequent implication of a constant average column density for individual GMCs in the Milky Way. 

We find that Milky Way wide measurements of the mass-size relation using \twco observations are well described by $M_{GMC} \sim R_{GMC}^2$ or a constant column density scaling relation. Two independent studies of the DHT Galactic \twco survey return nearly identical values of slopes, scaling coefficients and scatter characterizing the relation. The derived scaling coefficient corresponds to an average mass surface density of \sdsp $=$ 25 \sdu, for individual GMCs in the Milky Way. This value is considerably lower than that ($\sim$ 170 \sdu) typically found or assumed in the CO literature. In both studies the logarithmic scatter in the mass-size relation is quite large (0.45 dex). 

We find that two independent extinction based measurements of the mass-size relation of Milky Way GMCs are also well described by $M_{GMC} \sim R_{GMC}^2$ or a constant column density scaling relation. The two extinction studies are in excellent agreement with each other returning nearly identical slopes and scaling coefficients. However, both the scaling coefficient and scatter in the dust based relations differ from those found for the CO observations. The dust derived scaling coefficient corresponds to \sdsp $=$ 41.2 \sdu. The logarithmic scatter in the relations from the two extinction studies (0.04 \& 0.18 dex) is significantly lower than that of the CO observations.

We show that much of the difference in the scatter between the mass-size scaling relations derived from the dust and CO catalogs is due to a significant systematic variation of \sdsp with Galactic radius that is only apparent in the CO data because they cover a considerably larger range in Galactic radius than the extinction surveys.  We decompose this radial variation into two components. The first corresponds to an underlying linear gradient of unblended GMC surface densities which decreases with Galactocentric radius. The second component corresponds to a broad and strong peak in the surface density distribution above the linear gradient at the location of the well known molecular ring between 4-7 kpc from the center of the Galaxy. 

We find that the functional form of the linear unblended gradient is similar to that of the radial distribution of metallicity in the Milky Way. We posit that the unblended gradient in surface density is the result of the adoption of a constant rather than metallicity dependent CO mass conversion or X-factor by the two \twco catalogs. Our analysis of this surface density gradient suggests  that \xco varies inversely with metallicity. 
We derive an explicit expression for the radial dependence of X$_{CO}$ in the Milky Way. 

We suggest that the peak in \sdsp in the molecular ring may present the best evidence for a departure from  a constant GMC surface density in the Milky Way. However, we find that  cloud overlap and blending in the molecular ring produces an upward bias for measuring  GMC surface densities in that region. The size of this observed peak in \sdsp and the magnitude of the departure from a constant GMC surface density in the molecular ring could be much lower than implied by the measured values. 

The systematic difference in the scaling coefficients of the mass-size relation derived from CO and dust observations suggests that the X-factor  adopted by the CO studies (2 $\times$ 10$^{20}$ cm$^{-2}$ (K-km s$^{-1}$)$^{-1}$) underestimates the GMC masses. To minimize the effect of the radial variation in \sdsp we compare the mass-size relations of the two tracers for clouds located between 8-9 kpc from the center of the Galaxy. We derive the conversion factor for these GMCs to be X$_{CO}$ $=$ 3.6 $\pm$ 0.3 cm$^{-2}$ (K-km s$^{-1}$)$^{-1}$.  

We conclude that the bulk of the observed GMC population in the Milky Way can be described by a constant mass surface density of \sdsp $=$ 35 $\pm$ 8 \sdu.  The surface densities of GMCs in the molecular ring depart from  this constant surface density value but the size and nature of the departure are unclear at the present time.

\acknowledgements
{\it\noindent Acknowledgments\\}
We thank Professor B.-Q Chen for sending us results from the Chen et al. survey that were not included in the  published paper. We also thank Chris McKee, Leo Blitz and an anonymous referee for providing criticisms and suggestions that improved the paper. We acknowledge informative discussions with John Lewis and Catherine Zucker.

\clearpage

\end{document}